\documentclass[aps,prd,twocolumn,showpacs,preprintnumbers,superscriptaddress,dblfloatfix]{revtex4-1}
\usepackage{graphicx}   
\usepackage{dcolumn}    
\usepackage{bm}         
\usepackage{amssymb}    
\usepackage{setspace}
\usepackage{amsmath, amssymb, setspace}
\usepackage{array}
\usepackage{booktabs}
\usepackage{mathrsfs}
\usepackage{indentfirst}
\usepackage{slashed}
\usepackage{float}
\usepackage{lmodern}
\usepackage{hyperref}
\usepackage{xcolor}
\usepackage{multirow}
\usepackage{epstopdf}
\usepackage{ulem}
\usepackage{tabularx}
\usepackage[justification=raggedright]{caption}
\usepackage[flushleft]{threeparttable}
\usepackage{hyperref}
\usepackage{soul}
\usepackage[utf8]{inputenc}
\usepackage{notoccite}

\begin{document}

\preprint{IPMU18-0008}

\title{Primordial Black Holes from Inflaton Fragmentation into Oscillons}

\author{Eric Cotner}
\email[]{ecotner@physics.ucla.edu}
\affiliation{Department of Physics and Astronomy, University of California, Los Angeles\\
Los Angeles, CA 90095-1547, USA}

\author{Alexander Kusenko}
\email[]{kusenko@ucla.edu}
\affiliation{Department of Physics and Astronomy, University of California, Los Angeles\\
Los Angeles, CA 90095-1547, USA}
\affiliation{Kavli Institute for the Physics and Mathematics of the Universe (WPI), UTIAS\\
The University of Tokyo, Kashiwa, Chiba 277-8583, Japan}

\author{Volodymyr Takhistov}
\email[]{vtakhist@physics.ucla.edu}
\affiliation{Department of Physics and Astronomy, University of California, Los Angeles\\
Los Angeles, CA 90095-1547, USA}

\date{\today}

\begin{abstract}
We show that fragmentation  of the inflaton  into long-lived  spatially  localized  oscillon configurations  can lead to copious production of black holes. In simple inflation models  primordial black holes of {\it sublunar} mass can form, and they can account for all of the dark matter. We also explore the possibility that solar-mass  primordial  black holes,  particularly  relevant for gravitational wave astronomy, are  produced from the same mechanism.  
\end{abstract}

\maketitle

\section{Introduction}

Primordial black holes (PBHs) can form in the early Universe and can account for all or part of the dark matter (DM)~(e.g.~\cite{Zeldovich:1967,Hawking:1971ei,Carr:1974nx,GarciaBellido:1996qt,Khlopov:2008qy,Frampton:2010sw,Kawasaki:2016pql,Cotner:2016cvr,Carr:2016drx,Inomata:2016rbd,Inomata:2017okj,Garcia-Bellido:2017aan,Inoue:2017csr,Georg:2017mqk,Inomata:2017bwi,Kocsis:2017yty,Ando:2017veq}). They have also been linked to a variety of topics in astronomy, including the recently discovered \cite{Abbott:2016blz,Abbott:2016nmj,Abbott:2017vtc} gravitational waves \cite{Nakamura:1997sm,Clesse:2015wea,Bird:2016dcv,Raidal:2017mfl,Eroshenko:2016hmn,Sasaki:2016jop,Clesse:2016ajp,Takhistov:2017bpt}, formation of supermassive black holes~\cite{Bean:2002kx,Kawasaki:2012kn,Clesse:2015wea} as well as  $r$-process nucleosynthesis~\cite{Fuller:2017uyd}, gamma-ray bursts and microquasars \cite{Takhistov:2017nmt}  from compact star disruptions.

Many proposed scenarios of PBH formation assume that inflation has generated some excess of density perturbations on certain scales, which produce PBHs when they re-enter the Hubble horizon during the radiation dominated phase or during some intermediate matter-dominated stage (for review, see~\cite{Polnarev:1986bi,Khlopov:2008qy}).  The required inflaton potentials could be  {\it ad hoc}, or can be well-motivated in the context of hybrid inflation~\cite{GarciaBellido:1996qt}, supergravity~\cite{Kawasaki:2016pql}, etc. Recently, it has been shown that PBHs can also form from large extended objects associated with complex scalar fields, such as non-topological solitons in supersymmetric theories (Q-balls), which behave as matter and come to dominate the universe for a short time before decaying~\cite{Cotner:2016cvr,Cotner:2017tir}.  In this case, the overdensities needed for PBH formation result from statistical fluctuations in a system with a relatively low number of very massive ``particles" and not from the spectrum of primordial density fluctuations. In this work we show that if the inflaton potential admits long-lived oscillon solutions, associated with real scalar fields, their formation can lead to copious production of PBHs. While some recent work (e.g. \cite{Amin:2010xe}) have hinted at such possibility, we present the first explicit demonstration that this is a viable scenario. Further, we connect the results with a broader problem of dark matter. This opens a new avenue for formation of PBHs that constitute all of DM, while avoiding some stringent constraints associated with the
other production mechanisms in the context of inflationary models.

\section{Oscillon formation from the Inflaton}

Oscillons~\cite{Bogolyubsky:1976nx,Gleiser:1993pt,Copeland:1995fq,Fodor:2006zs,Kasuya:2002zs,Honda:2001xg} arise in many well motivated theories with scalar fields, such as models of inflation \cite{Amin:2011hj,Hong:2017ooe}, axions \cite{Kolb:1993hw} or moduli~\cite{Antusch:2017flz}.  The oscillons are localized, metastable, pseudo-solitonic configurations of real scalar fields. The stability of an oscillon is not guaranteed by a  conserved charge and its long lifetime is associated with an  approximate adiabatic invariant~\cite{Kawasaki:2015vga,Kasuya:2002zs}. 
Early Universe oscillons have been recently studied in connection with primordial gravity waves \cite{Antusch:2016con} as well as baryogenesis~\cite{Lozanov:2014zfa}.  

For definiteness, we consider the model with a single inflaton field $\phi$ that has a canonical kinetic term, minimal coupling to Einstein gravity and a potential~\cite{Amin:2010xe,Amin:2010dc,Amin:2011hj,Amin:2010jq}
\begin{equation} \label{eq:model}
V(\phi) = \dfrac{m^2}{2} \phi^2 -  \dfrac{\lambda}{4} \phi^4 + \dfrac{g^2}{6 m^2} \phi^6~.
\end{equation}
Here $\lambda > 0$ and for convenience, following the original studies, we take $\lambda, g, m/ M_{\rm pl} \ll 1$, where $M_{\rm pl}$ is the Planck mass.
The model~\cite{Amin:2010xe,Amin:2010dc,Amin:2011hj,Amin:2010jq} is inspired by supergravity and string theories~\cite{McAllister:2008hb,Silverstein:2008sg,Amin:2011hj}, and the potential could be considered as a Taylor series expansion of a more general potential for some range of the scalar field.~We assume that density perturbations that seed cosmological structures are generated when the inflaton field has a much larger value, for which the shape of the potential is not necessarily described by Eq.~(\ref{eq:model}).~After the inflationary phase, the inflaton begins to oscillate near the minimum of the potential as described by Eq.~(\ref{eq:model}).  At this time, the inflaton condensate fragments into oscillons. Hence, the details of inflationary physics are somewhat tangential to oscillon formation. Later in the text we present an example of a viable inflationary model that can accommodate this potential.

A necessary condition for oscillon formation is that the potential is shallower than quadratic near the minimum (making the scalar self-interactions attractive). This is the case for $\lambda>0$. 
An initially homogeneous inflaton condensate can fragment into oscillon lumps after the inflaton's self-resonance parametrically amplifies field fluctuations $\delta \phi_k$ in some band of wave-numbers $k$ around the background field $\overline{\phi}$. This can be analytically investigated through the Floquet analysis.
For $(\lambda / g)^2 \ll 1$, the above potential admits ``flat-top'' oscillons,  on which we focus in our analysis.  They are extremely stable on the cosmological time scales, and they admit analytic description~\cite{Amin:2010xe,Amin:2010dc,Amin:2011hj,Amin:2010jq}. 

We summarize the main features of oscillon formation below.
An initially homogeneous inflaton condensate can fragment into lumps, corresponding to oscillons. The inflaton self-resonance parametrically amplifies field fluctuations $\delta \phi_k$ in some band of wave-numbers $k$ around the background field $\overline{\phi}$. This can be analytically investigated through Floquet analysis, where the most unstable modes behave as 
\begin{equation} 
\delta \phi_k (t) \propto e^{\mu_k t} P(t)~,
\end{equation}
with $\mu_k$ denoting the Floquet exponent and $P(t)$ a periodic function. In an expanding background significant amplification of fluctuations requires $\mu_k (a) / H \gg 1$, where $a(t)$ is the cosmic scale factor and $H = H_i / \sqrt{a^3}$ is the Hubble parameter at the bottom of potential.~At fragmentation $H_i \simeq \sqrt{\lambda/10 g^2} (m/M_{\rm pl}) m$ and $a_i = 1$. The amplification condition translates to~\cite{Amin:2010xe} 
\begin{equation}
\dfrac{\mu_k (a)}{H} = \dfrac{M_{\rm pl}}{m} \Big( \dfrac{\lambda^{3/2}}{g}\Big) \Big[ \sqrt{\dfrac{9}{4} \dfrac{\tilde{k}^2}{a^2} \Big( 1 - \dfrac{1}{a^3}\Big) - \Big(\dfrac{\tilde{k}^4}{a}\Big)}\Big] \gg 1~,
\end{equation}
where $\tilde{k} = (g/\lambda m) k$, with $k$ being related to the physical wavenumber via $k = a \, k_p$. The total amplification of fluctuations as they pass through the instability band is found by integrating over the Floquet exponent as 
\begin{equation}
\delta \phi_k (a) \sim \dfrac{1}{\sqrt{2 \omega_k}} \dfrac{1}{a^{3/2}} e^{\beta f(\tilde{k},a)}~,
\end{equation}
where 
\begin{equation}
f(\tilde{k},a) = \sqrt{\dfrac{5}{2}} \int_{C} d \log{\overline{a}} \Big[ \tilde{k} \sqrt{\dfrac{9}{10 \overline{a}^2} \Big(1 - \dfrac{1}{\overline{a}^3}\Big) - \dfrac{\tilde{k}^2}{\overline{a}}} \Big]
\end{equation}
and $C = \frac{a^3 -1}{a^4} > \frac{10}{9} \tilde{k}^2$, $\beta = \sqrt{\lambda} (\lambda/g) (M_{\rm pl}/m)$ and $\omega_k^2 \simeq k^2 + m^2$. Since $\beta \sim \mu/H$, we are interested in the $\beta \gg 1$ regime.

The condition~\cite{Amin:2010xe} for formation of oscillons from amplified perturbations can be formulated as $k^{3/2} \delta \phi \sim \overline{\phi}$. 
The average number density of oscillons can then be estimated as
\begin{equation}
\overline{n} \sim (k_{nl}/2 \pi)^3 / a^3~,
\end{equation}
where $k_{nl} \sim \beta^{-1/5} (\lambda/g) m$
label the modes that become non-linear the earliest. While the model supports several distinct oscillon populations, we focus on the flat-top oscillons due to their stability.~Unlike the usual Gaussian-profile oscillons, they possess an approximately uniform core density of 
$\rho_c \simeq m^4 (9 \lambda/20 g^2)$~\cite{Amin:2010dc}. 
 Taking the characteristic radius of oscillons to be $R \sim \pi/k_{nl}$, we can  estimate their energy as
\begin{equation}
E \sim \dfrac{4 \pi \rho_c R^3}{3} \simeq \dfrac{3 \pi^4 m^4}{5 k_{nl}^3} \Big(\dfrac{\lambda}{g^2}\Big)~.
\end{equation}
The above provides $\overline{n}(E)$ through $k_{nl}$ substitution.

\section{Statistical Fluctuations of the Field Lumps}

Probability distribution of oscillons can be obtained from statistical fluctuations.  
Given an average number density $\overline{n}$ of uniformly distributed objects, the probability of finding $N$ objects within a volume $V$ follows the Poisson distribution
\begin{equation}
P_N (N) = \dfrac{(\overline{n} V)^N}{N!} e^{- \overline{n} V}~.
\end{equation}
 The Poisson distribution originates from binomial distribution of independent random events in the large event limit. Thus, this choice is the most general description assuming that fragmentation events are uncorrelated and allows for analytic treatment. A deviation from this distribution would be highly model dependent.  
 
From above, one can derive the probability distribution $P_{\eta} (\eta)$ that corresponds to number density of oscillons of reference energy $E_0$, related to $E$ via $E = E_0 / \tilde{k}_{nl}^3$, where $\eta = M/E_0$ and \begin{equation}
E_0 = \Big(\dfrac{3 \pi^4}{5}\Big) \Big[\Big(\dfrac{M_{\rm pl}}{\beta^2}\Big)\Big(\dfrac{\lambda}{g}\Big) \Big(\dfrac{M_{\rm pl}}{m}\Big)  \Big]~.
\end{equation}
The total mass of a cluster of oscillons is $M = N E$. Hence, the probability distribution of oscillon cluster mass is given by 
\begin{equation}
P_M (M) = \sum_{N} P_N (N) \delta (M - NE)~.
\end{equation}
The delta-function can be eliminated through a Fourier transform
\begin{equation}
\tilde{P}_M (\mu) = \int dM \, P_M (M) \, e^{i M \mu} = e^{\overline{n} V (e^{i E \mu} -1)}~,
\end{equation}
followed by an inverse transform
\begin{equation} \label{eq:pmint}
P_M (M) = \dfrac{1}{2 \pi} \int d\mu  \, e^{-i M \mu} \, e^{\overline{n} V (e^{i E \mu} -1)}~.
\end{equation}
An approximate analytic non-integral form of  Eq. \eqref{eq:pmint} can be found through the method of steepest descent. The resulting expression is 
\begin{equation}
P_{\eta} (\eta) = \dfrac{1}{\sqrt{2 \pi \beta^{3/5} \eta}} e^{\beta^{-3/5}[\eta (1 - \ln\{(2 \pi)^3 \eta/v \}) - v/(2 \pi)^3]}~,
\end{equation}
where 
\begin{equation}
v = V \Big(\dfrac{\lambda m}{g}\Big)^3 = V \Big[M_{\rm pl} \Big(\dfrac{\lambda}{g}\Big) \Big(\dfrac{m}{M_{\rm pl}}\Big)\Big]^3
\end{equation}
and $\eta = M/E_0$ denote  the rescaled dimensionless volume and mass, respectively.

\section{Black Hole Formation}

Using $P_{\eta} (\eta)$ we can now calculate the distribution of the initial density contrasts $\delta_0$, $P_{\delta_0}(\delta_0)$. In terms of $\eta$, $\delta_0$ is
\begin{equation} \label{eq:density}
\delta_0 = \dfrac{\delta \rho}{\overline{\rho}} = \dfrac{\rho - \overline{\rho}}{\overline{\rho}} = \dfrac{M/V - \overline{\rho}}{\overline{\rho}} = (2\pi)^3 \Big( \dfrac{\eta}{v}\Big) - 1~,
\end{equation}
where 
\begin{equation}
\overline{\rho} = \overline{n} \, E  = \dfrac{1}{(2\pi)^3} E_0 \Big(\dfrac{\lambda m}{g}\Big)^3
\end{equation}
denotes the average background energy density of oscillons, with the scale factor in $\overline{n}$ evaluated at $a = 1$.

\begin{figure}[t]
\centering
\hspace{-2em}
\includegraphics[trim = 0.0mm 0.0mm 0.0mm 0mm, clip,width = 3in]{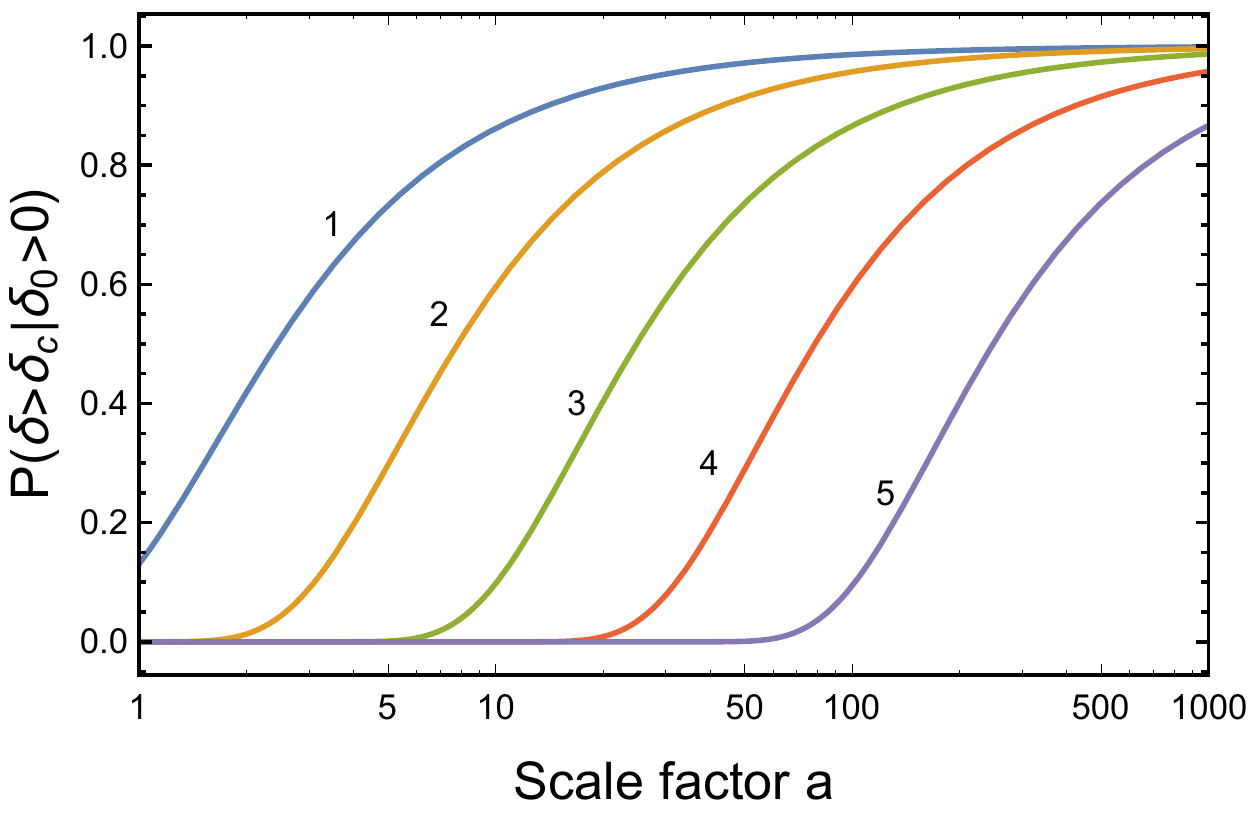}\\
\caption[]{Fraction of super-critical overdensities as a function of the scale factor $a(t)$. Results are shown for several different volumes, which are expressed in terms of the Hubble horizon as $V/V_{H} = 3 \times 10^{-5}$ , $3 \times 10^{-4}$, $3 \times 10^{-3}$ , $3 \times 10^{-2}$ , $3 \times 10^{-1}$ and labeled ``1'', ``2'', ``3'', ``4'' and ``5'', respectively. The values of the input parameters $(\lambda/g)^2 = 0.2$ and $\beta = 50.4$ correspond to those of Model A.}
\vspace{-1em}
\label{fig:pdeltacrit}
\end{figure}

The initial overdensities evolve due to gravitational self-attraction and grow according to the scale factor during the oscillon matter-dominated era as $\delta(t) = \delta_0 a(t) = \delta_0 (t/t_0)^{2/3}$.~Once overdensities $\delta$ exceed the critical threshold $\delta_c \sim 1$,  regions start collapsing and forming black holes.
The total fraction of super-critical overdensities as a function of scale factor $a$ can be found through integration
\begin{equation}
P(\delta \geq \delta_c) = \int_{\delta_c = 1}^{\infty} \frac{d\delta}{a}P_{\delta_0}(\delta/a)~.
\end{equation}
In Figure~\ref{fig:pdeltacrit} we display this fraction for several different volume values.~As can be seen, already within a few $a$ from the time of fragmentation the amount of super-critical regions becomes significant.

Not all super-critical regions result in a black hole. 
Unlike the radiation-era PBH formation \cite{Carr:1975qj}, absence of pressure gradient in the matter-dominated era greatly enhances black hole production \cite{Polnarev:1986bi}. On the other hand, non-spherical density anisotropies now play a dominant role with final stages of collapse being described by a ``Zel'dovich pancake'' \cite{Zeldovich:pancake}, whose parameter distribution can found in \cite{Doroshkevich:collapse}. 
Using Thorne's hoop conjecture \cite{Thorne:1972ji} as a requirement for formation of the black hole horizon, PBH production was recently re-analyzed in \cite{Harada:2016mhb} (see discussion in text for comparison with \cite{Polnarev:1986bi}). 
The probability for a super-critical overdensity region to result in a black hole is given by
  \begin{equation}
    B(M) \simeq
      0.056 \, \delta_0^5 \left(\dfrac{M}{\overline{M}(V_H)}\right)^{10/3}~,
  \end{equation}
where $\overline{M}(V_H) =  (4 \pi/3) \overline{\rho} / H_i^3$  denotes the average mass in the Hubble horizon volume $V_H(t) = (4 \pi/3) t^3$ at fragmentation, with $t = 1/H$. We have further checked that including the effects of PBH spins \cite{Harada:2017fjm}, relevant for small overdensities, will not significantly alter our results.~The probability of an overdensity region to form a BH controls the amount of energy budget that remains available for reheating the Universe.

\begin{figure*}[ht]
\begin{minipage}[b]{0.47\textwidth}
\centering
\includegraphics[width=\linewidth]{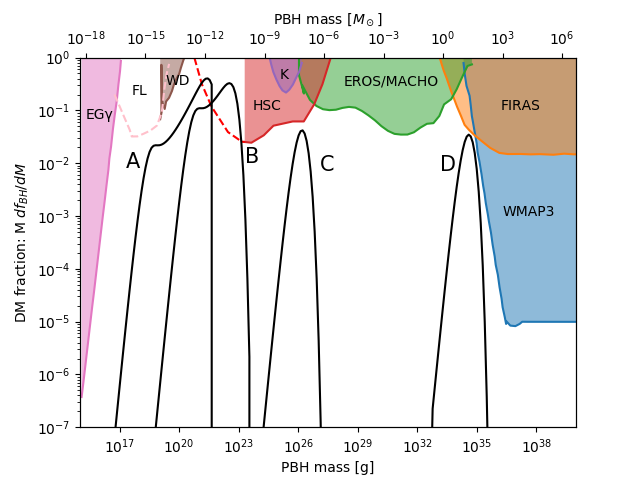}
\vspace{-2.85mm}
\end{minipage}
\begin{minipage}[b]{0.475\textwidth}
\centering
\includegraphics[width=\linewidth]{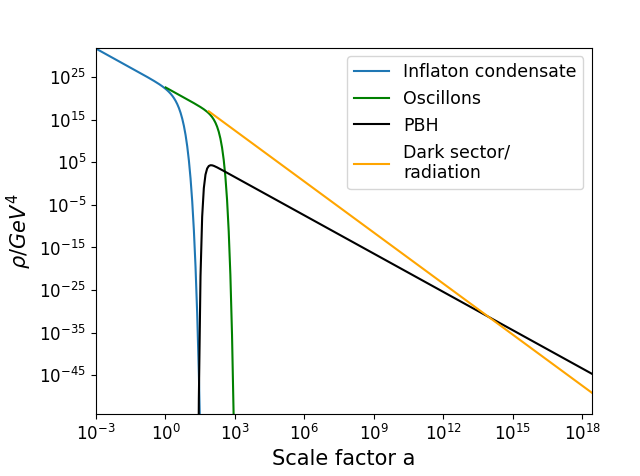}
\end{minipage}
\caption[]{[Left] DM fraction in primordial black holes. Fits for parameter choices corresponding to Model A, B, C and D are shown.
Constraints from extragalactic $\gamma$-rays from BH  evaporation~\cite{Carr:2009jm} (EG$\gamma$), femto-lensing~\cite{Barnacka:2012bm} (FL), white dwarf abundance~\cite{Graham:2015apa} (WD), Kepler star milli/micro- lensing~\cite{Griest:2013aaa} (K), Subaru HSC micro-lensing~\cite{Niikura:2017zjd} (HSC) and MACHO/EROS/ OGLE micro-lensing~\cite{Tisserand:2006zx} (ML) are displayed. Dashed FL line
indicates that extended nature of the source as well as wave optics effects render GRB femtolensing constraints less effective \cite{Katz:2018zrn}.
Dashed HSC line indicates that 
constraints are expected to be weaker than reported when PBH Schwarzschild radius becomes smaller than
the wavelength of light~\cite{Takahashi:2003ix,Matsunaga:2006uc}. [Right] Evolution of cosmological density for parameters of Model A. Contributions from the un-fragmented inflaton, oscillons, PBHs as well as the radiation sector are shown.
}
\label{fig:PBHcosmo}
\end{figure*}

\begin{table*}[ht]
\setlength{\extrarowheight}{3pt}
  \setlength{\tabcolsep}{7pt}
  \begin{center}
  \begin{threeparttable}
	\begin{tabular}{|c|c|c|c|c||c|c|c|c|c|}  \hline 
	 Model & $\beta$ & $(\lambda/g)^2$ & $a_R$   & $m_{\phi}$  
    & $\overline{\phi}_{\rm frag}$  
   & $H_i$   & $\Gamma_{\phi}$ & $T_{\rm R}$   & $f_{\rm PBH}$ \\
   	 &   &   &     &   (GeV)
    &   (GeV)
   &   (GeV) &  (GeV) &  (GeV) &   \\ \hline \hline
   	A & 50.4 & 0.2 & 70 & $9 \times 10^{-6}$  
    & $7.5 \times 10^{15}$
   & $9.2 \times 10^{-8}$   & $2.4 \times 10^{-11}$ & $3.0 \times 10^3$   & $1.0$ \\
        \hline
   	B & 35.5 & 0.2 & 20 & $9 \times 10^{-8}$  
    & $1.1 \times 10^{16}$
   & $1.3 \times 10^{-9}$   & $2.2 \times 10^{-12}$ & $9.0 \times 10^2$   & $1.0$ \\
        \hline 
        
           	C & 30.0 & 0.3 & 2 & $6 \times 10^{-13}$  
    & $1.9 \times 10^{16}$
   & $1.5 \times 10^{-14}$   & $8.2 \times 10^{-16}$ & $1.7 \times 10^{1}$   & $6.2 \times 10^{-2}$ \\
   
        \hline         
   	D & 10.7 & 0.2 & 2 & $1 \times 10^{-20}$  
    & $5.3 \times 10^{16}$
   & $7.2 \times 10^{-22}$   & $3.8 \times 10^{-23}$ & $3.8 \times 10^{-3}$   & $5.1 \times 10^{-2}$ \\
        \hline 
 	\end{tabular}
\caption{\label{tab:param} Parameter sets for three specific model realizations (Model A, B, C, D). In model A/B PBHs can account for all of the DM, model C corresponds to the region where HSC experiment \cite{Niikura:2017zjd} has reported a candidate PBH event, while model D allows for PBHs to contribute to the observed LIGO black hole merger events. Vertical double line divides the input quantities [left-side] $\beta$, $(\lambda/g)^2$, reheating time $a_R=a(T_R)$, inflaton mass $m_{\phi}$ and the derived quantities [right-side]: inflaton VEV at fragmentation $\overline{\phi}_{\rm frag}$, initial Hubble rate $H_i$, inflaton decay rate $\Gamma_{\phi}$, reheating temperature $T_R$ and the fraction of DM in PBHs $f_{\rm PBH}$.}
\label{tab:experiments}
\end{threeparttable}
  \end{center}
\end{table*}

At some scale factor $a_R$ the oscillon matter-dominated era ends and the Universe is reheated, entering the radiation-dominated phase. Specifying the energy density of the overdensities as $\rho = M/V = E_0 \eta /V$, the PBH spectrum at $a_R$ is given by
\begin{align}
\dfrac{d \langle \rho_{\rm PBH} \rangle}{d\eta} = &\int_{V_{\rm min}}^{V_{\rm max}} \dfrac{dV}{V}\left(B(\eta) \dfrac{E_0 \eta}{V}\right) P_{\eta}(\eta) \notag\\
& \times \theta[\eta - \eta_c (a_R, V)] \theta\left[ \rho_0 - \dfrac{E_0 \eta}{V}\right]~,
\end{align}
where $\eta_c$ is the critical mass, $\theta[x]$ is the Heaviside step function and $V_{\rm min}$, $V_{\rm max}$ are the average volume of a single oscillon and the Hubble horizon volume, respectively.~The first step function selects super-critical regions.~The second step function imposes energy conservation by requiring that an overdensity doesn't exceed the inflaton energy density, assuming that both have the same volume. The inflaton energy density $\rho_0$ at the bottom of its potential can be found from the mass term $\rho_0 a_i^3 = (1/2) m^2 \overline{\phi}_i$, where $\overline{\phi}_i= \sqrt{3 \lambda/5 g^2} m$. A similar relation can be obtained directly from the Friedmann equations.
In order to get the present day distribution we must redshift the results obtained at fragmentation time. The redshift factor 
\begin{equation}
\Big(\dfrac{a_F}{a_R}\Big) = \dfrac{g_{\ast S}^{1/3}(T_F) T_F}{g_{\ast S}^{1/3}(T_R) T_R}
\end{equation}
accounts for evolution from $T_R$ (defined by $\rho_R (T_R) = (\pi^2/30) g_{\ast} (T_R) T_R^4$) to $T_0 = 2.7$ K = 2.3 meV. Here, $g_{\ast}$ denotes the relevant number of relativistic degrees of freedom.
Reverting from $\eta$ back to $M$, the fraction of DM residing in PBHs is then 
\begin{equation}
\dfrac{d f_{\rm PBH}}{dM} =  \dfrac{1}{\rho_{\rm DM} a^3}  \dfrac{d \langle \rho_{\rm PBH} \rangle}{dM}~,
\end{equation}
where $\rho_{\rm DM}$ is the present-day DM density and the $1/a^3$ factor accounts for the redshift.  In Figure \ref{fig:PBHcosmo} we display the fraction of PBHs as DM for several specific parameter sets (denoted as ``Model A'', ``B'', ``C'', ``D'') along with the current experimental constraints.~The constraints shown are valid only for monochromatic mass functions, but can be adapted for extended mass functions by following equations (12) and (13) of Ref.~\cite{Carr:2017jsz}.~We have verified numerically that these extended mass function constraints are satisfied.~We note that, in the mass ranges of $10^{17} - 10^{19}$ g and $10^{20}-10^{23}$~g, PBHs can account for all of the dark matter.~The extended nature of the sources as well as wave optics
effects render GRB femtolensing constraints around PBH
mass of $10^{18}$ g ineffective~\cite{Katz:2018zrn}.~The HSC constraints reported in Ref.~\cite{Niikura:2017zjd} do not apply below $10^{23}$~g because the micro-lensing magnification is strongly suppressed when the Schwarzschild radius of the black hole becomes smaller than the wavelength of light~\cite{Takahashi:2003ix,Matsunaga:2006uc}.
In the same mass range, one could use the stability of neutron stars to constrain PBHs if globular clusters contained $\gtrsim 10^3$ times the average dark matter density~\cite{Capela:2013yf}. However,
observations of globular clusters show no evidence of  dark matter content in such systems, resulting in upper bounds three order of magnitude below the levels needed to allow for meaningful constraints~\cite{Bradford:2011aq,Sollima:2011nb}. 

In the above, model A/B correspond to the region where PBHs can make up all of the DM, while model D covers the region where PBHs can contribute to the observed LIGO black hole merger events \cite{Bird:2016dcv, Sasaki:2016jop,Clesse:2016ajp}. The intermediate case between models A/B and model D, model C, corresponds to the region where HSC reported one PBH candidate event  \cite{Niikura:2017zjd}. Exact values of the parameters, including both the input and the derived quantities, can be found in Table \ref{tab:param}. We note that Model D is phenomenologically not viable and the relevant parameters are shown for completeness.

\section{Cosmological Aspects}

\subsection{Reheating}

In Figure \ref{fig:PBHcosmo} we display the cosmological history of the setup, showing energy density evolution of the inflaton, oscillons, PBHs as well as radiation from reheating. During inflation, the inflaton  dominates the Universe. As the inflaton settles at the bottom of potential the Universe becomes matter-dominated with the density scaling as $a^{-3}$. After fragmentation of the inflaton into oscillons and PBH formation the Universe is reheated, becoming radiation-dominated with the density scaling as $a^{-4}$. At redshift $z \approx 3600$ dark matter in the form of PBHs, whose density scales also as $a^{-3}$, overtakes the radiation contribution, and the Universe again enters matter-dominated regime. Unlike the case of Q-balls~\cite{Cotner:2017tir,Cotner:2016cvr}, there is no intermediate radiation-dominated era before the fragmentation time in our setup, since oscillons form directly from the inflaton during the early stages of reheating.

The Universe is reheated from oscillon decay.  If PBHs are to constitute a significant fraction of DM, the inflaton must be very light, and the reheating temperature is very low (model A/B, Table \ref{tab:param}) compared to the typically considered values for thermal cosmology. To avoid affecting the Big Bang nucleosynthesis (BBN), reheating should occur above $ T \gtrsim 4$ MeV scale (e.g. \cite{deSalas:2015glj}).  While it is commonly assumed that the Universe was reheated to a much higher temperature, a cosmological history with $\sim$MeV reheating is possible and is consistent with observations~\cite{Yaguna:2007wi,Gelmini:2008fq}.   Neglecting the oscillon quantum decay \cite{Hertzberg:2010yz}, the allowed direct inflaton decay channels are limited by the total invariant mass. The simplest decay mode is to photons $\phi \rightarrow \gamma\gamma$, proceeding through an effective $(g_{\gamma\gamma}/4) F_{\mu \nu} F^{\mu \nu} \phi$ operator, where $F_{\mu \nu}$ is the electromagnetic field strength tensor and $g_{\gamma\gamma}$ is the coupling. The relevant decay rate is given by 
\begin{equation}
\Gamma_{\phi \rightarrow \gamma\gamma} = (g_{\gamma \gamma}^2/64 \pi) m_{\phi}^3~.
\end{equation}
However, axion-like particle searches already strongly constrain this channel \cite{Jaeckel:2017tud,Patrignani:2016xqp}.  Thus, without an extended dark sector, Model D is not viable. In a more complicated model, the reheating may be possible if the inflaton decays into some dark sector particles, which produce the Standard Model degrees of freedom via mixing (e.g. \cite{Tenkanen:2016jic,Adshead:2016xxj}). Generating the matter-antimatter asymmetry in a low-reheating scenario also presents a model building challenge.

\subsection{A Viable Inflationary Model}

Taken at face value, the potential of Eq. \eqref{eq:model} produces unphysical perturbation spectrum during inflation due to the dominance of $\phi^6$ term (see \cite{Amin:2011hj} for discussion). However, the field value $\overline{\phi}$ at the bottom of potential is far below the Planck scale that sets the initial inflaton displacement. Hence, our region of interest where inflaton oscillates near the potential minimum is decoupled from the large values that determine the inflationary phase. In this work we remain agnostic regarding the exact shape of the potential and the early Universe dynamics.  We focus on the effective potential at a relatively small vacuum expectation value (VEV), well below the values that are relevant for structure formation.  

Below, we present an example of a more general model that is also viable during the inflationary phase and that allows for oscillon PBH formation to occur. The original inflaton potential of Eq. (1) in the main text results in an unphysical curvature perturbation spectrum, dominated by the $\phi^6$ term. 
The problem can be alleviated if the potential is modified such that it leads to physical inflationary scenario at large $\phi$ field values, but reduces to the original form at the small field values, ensuring that oscillons form. 

This task can be accomplished by adding an extra field $\chi$, resulting in a 2-field model of the form
\begin{equation}
V(\phi, \chi) ~=~\dfrac{m^2}{2} \phi^2 -  \dfrac{\lambda}{4} \phi^4 + \dfrac{g^2}{6 m^2} \phi^6 + \dfrac{m_{\chi}^2}{2} \chi^2~,
\end{equation}
where the first three terms in the above equation constitute the original potential. The general idea is that if the initial field configuration has a large value of $\chi$, then inflation proceeds along the $\chi$ direction until it approaches the bottom of the potential, at which point the $\phi$ field takes over and produces oscillons. 

From the Friedmann equations the Hubble expansion rate is
\begin{equation}
H^2 ~=~ \dfrac{8 \pi}{3 M_{\rm pl}^2} \Big[\dfrac{1}{2}(\dot{\phi}^2 + \dot{\chi}^2) + V(\phi, \chi) \Big]~,
\end{equation}
with over-dots indicating derivatives with respect to cosmological time $t$.
The dynamics of inflation are described by the Klein-Gordon equation in the background of an expanding spacetime. For our 2-field model, the dynamics are described by 

\begin{equation}
  \begin{cases}
    \ddot{\phi} + 3 H \dot{\phi} + \dfrac{\partial{V(\phi, \chi)}}{\partial{\phi}} ~=~ 0\\
    \ddot{\chi} + 3 H \dot{\chi} + \dfrac{\partial{V(\phi, \chi)}}{\partial{\chi}} ~=~ 0~.
  \end{cases}
\end{equation}

\begin{figure}[t]
\centering
\vspace{2em}
\hspace{-2em}
\includegraphics[trim = 0.0mm 0.0mm 0.0mm 0mm, clip,width = 3in]{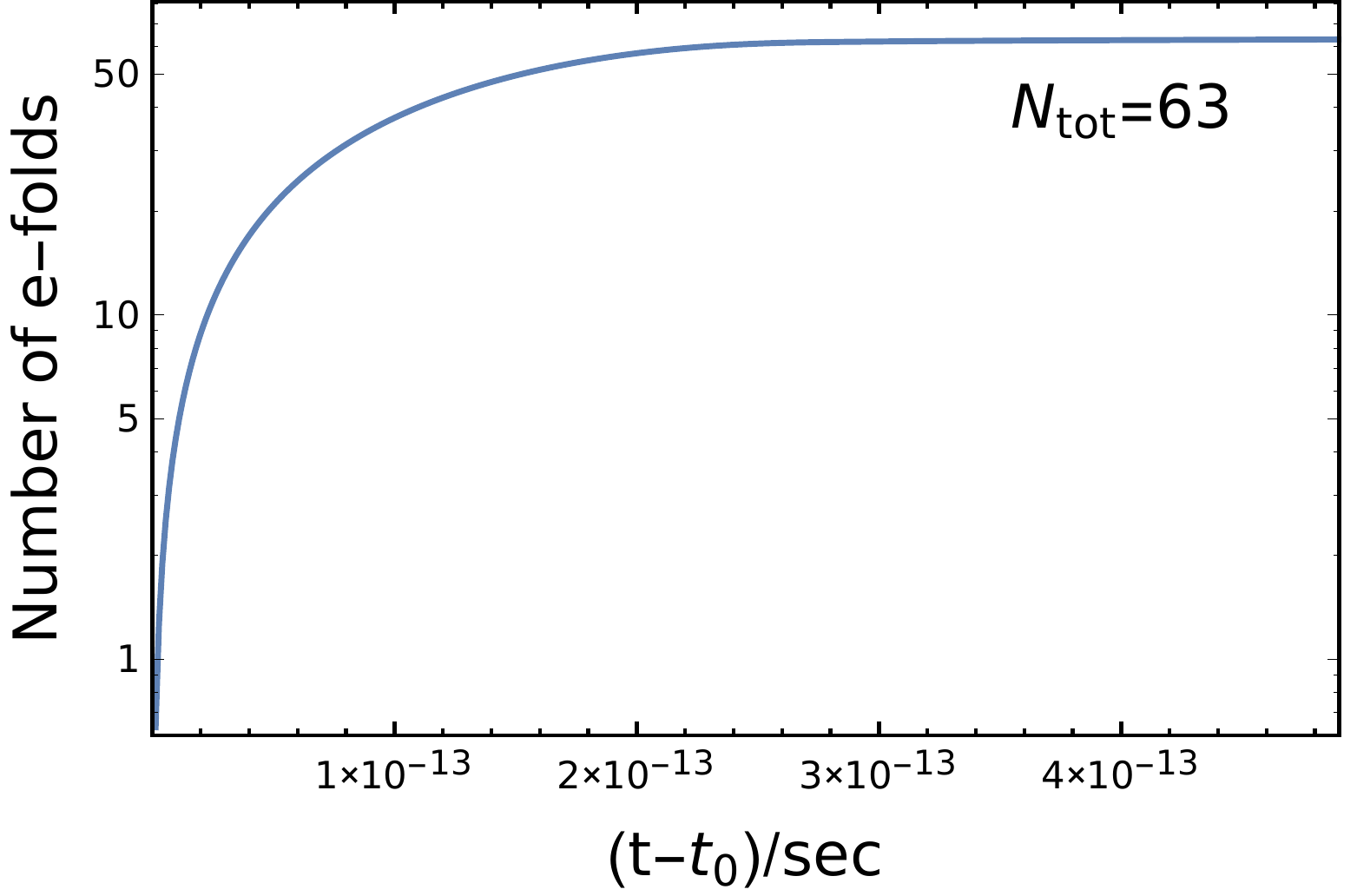}\\
\caption{Number of inflationary e-folds vs. time for the 2-field model. Input parameters for Model A from the main text are used.}
\label{fig:efolds}
\end{figure}

We set the model parameters to correspond to our Model A in Table 1, allowing for PBHs to constitute all of the DM. For the $\chi$-field, we set $m_{\chi} = 6 \times 10^{-6} M_{\rm pl}$.
To solve the equations we specify the initial conditions at $t = 0$ as $\dot{\phi}(0) = \dot{\chi}(0) = 0$  and $\phi(0) = \overline{\phi}$, $\chi(0) = 5.1 \times 10^5 \, m_{\chi}$. Since $m_{\chi} \gg m$, the inflationary dynamics are dominated by the $\chi$ field and the scenario is well-approximated by single-field inflation. We have confirmed that cosmological evolution of $H$ is reasonable.~The resulting number of the inflationary e-folds $N = \int H dt$, depicted on Fig.~\ref{fig:efolds}, is observed to saturate at $N \sim 63$.

We can now compute the primordial curvature perturbation power spectrum $P_{\zeta}(k)$. Since the dynamics of inflation are dominated by the $\chi$ field, we can ignore the $\phi$ field and calculate the power spectrum directly from the slow-roll approximation. Far from the origin, the potential is approximately $V \simeq (1/2) m_{\chi}^2 \chi^2$. The respective amplitude of the curvature perturbations is
\begin{equation}
\Delta_{\zeta}^2 (k) = \dfrac{k^3}{2 \pi^2} P_{\zeta}(k) = \dfrac{1}{96 \pi^2} \Big(\dfrac{m_{\chi}}{M_{\rm pl}}\Big)^2 (4 N)^2~.
\end{equation}
For the number of e-folds obtained above, the experimental bound from \textit{Planck} 2018 of $\Delta_{\zeta}^2(k) = 2.105 \times 10^{-9}$~\cite{Aghanim:2018eyx} requires $m_{\chi} \lesssim 6 \times 10^{-6} M_{\rm pl}$, which is satisfied in our model.
~\newline
~\newline

\section{Summary}

We have shown that inflaton fragmentation into oscillons can lead to formation of primordial black holes in a single-field inflation model or other models that admit oscillon solutions. This novel production mechanism can generate a sufficient density of PBHs to account for all or part of dark matter.~It is also possible that solar-mass black holes can be produced this way, but the required mass of the inflaton is very small, and the need for reheating and baryogenesis will lead to more complicated models. 

\section*{Acknowledgements}

This work was supported by the U.S. Department of Energy Grant No. DE-SC0009937  A.K. was also supported by the World Premier International Research Center Initiative (WPI), MEXT, Japan.

\bibliography{pbhosc}
\end{document}